\documentclass[11pt]{article}
\pdfoutput=1
\usepackage[ascii]{inputenc}
\usepackage{hyperref,dcolumn,bbm,url,booktabs,braket,microtype,aliascnt,mathtools,mathrsfs,etoolbox,cancel,capt-of,algpseudocode,float,newfloat,amsmath,amssymb,color,amsthm,fullpage,cleveref}
\usepackage{subcaption}

\usepackage[T1]{fontenc}
\usepackage[USenglish]{babel}
\usepackage{ marvosym }
\usepackage{tabularx}
\usepackage{graphicx}
\usepackage{array, ltablex, multirow}
\usepackage{fancyvrb}

\crefname{thm}{theorem}{theorems}
\crefname{lem}{lemma}{lemmas}
\crefname{cor}{corollary}{corollaries}
\crefname{prb}{problem}{problems}
\crefname{dfn}{definition}{definitions}

\newcommand{\ba}{\begin{eqnarray}}
\newcommand{\ea}{\end{eqnarray}}
\newcommand{\TV}{\text{TV}}

\DeclareMathOperator{\diag}{diag}

\title{Hierarchical Learning for Quantum ML: Novel Training Technique for Large-Scale Variational Quantum Circuits}
\author{Hrant Gharibyan$^{1,2}$, Vincent Su$^{1,3}$, Hayk Tepanyan$^1$ \\  
\\ {\it $^1$BlueQubit Inc, 
Los Angeles, CA 90046, USA}
\\ {\it $^2$ Institute for Quantum Information and Matter, Caltech, Pasadena, CA 91125, USA}
 \\ {\it $^3$Department of Physics, University of California, Berkeley, CA 94720, USA}}

\begin{document}

\maketitle

\begin{abstract}
% In this article, we present numerical Quantum AI/ML experiments using a specific model of variational quantum circuits, called Quantum Circuit Born Machines (QCBM). This model have been studied for distirbution loading taks onto a qubit and have shown some promise for data loading. Here we propose a new architecure of generatize QCBM model, called Recursive QCBM (RQCBM) that is faster to train on the data and has higher accuracy then traditional QCBM models.
We present hierarchical learning, a novel variational architecture for efficient training of large-scale variational quantum circuits. We test and benchmark our technique for distribution loading with quantum circuit born machines (QCBMs). 
With QCBMs, probability distributions are loaded into the squared amplitudes of computational basis vectors represented by bitstrings.
Our key insight is to take advantage of the fact that the most significant (qu)bits have a greater effect on the final distribution and can be learned first.
One can think of it as a generalization of layerwise learning, where some parameters of the variational circuit are learned first to prevent the phenomena of barren plateaus.
We briefly review adjoint methods for computing the gradient, in particular for loss functions that are not expectation values of observables.
We first compare the role of connectivity in the variational ansatz for the task of loading a Gaussian distribution on nine qubits, finding that 2D connectivity greatly outperforms qubits arranged on a line.
Based on our observations, we then implement this strategy on large-scale numerical experiments with GPUs, training a QCBM to reproduce a 3-dimensional multivariate Gaussian distribution on 27 qubits up to $\sim4\%$ total variation distance.
Though barren plateau arguments do not strictly apply here due to the objective function not being tied to an observable, this is to our knowledge the first practical demonstration of variational learning on large numbers of qubits.
We also demonstrate hierarchical learning as a resource-efficient way to load distributions for existing quantum hardware (IBM's 7 and 27 qubit devices) in tandem with Fire Opal optimizations.

\end{abstract}

\newpage
\tableofcontents

\section{Introduction}

% Variational quantum circuits are a general class of circuits whose elements can be tweaked by some control parameter such as pulse duration or rotation angle. These parameters are then typically optimized via gradient descent to minimize certain quantities, such as the energy of a Hamiltonian. In the era of NISQ devices, a common subclass is known as the Hardware Efficient Ansatz (HEA)~\cite{kandala2017hardware}, where the gate elements are based on the native operations to the quantum hardware, reducing the overhead of \textit{compilation}. Since these operations are noisy, one also typically restricts to a relatively shallow circuit. From a learning perspective, this is a sweet spot~\cite{leone2022practical} in that deeper HEAs will suffer from barren plateaus, the phenomenon of the concentration of gradients around exponentially small values.

Quantum hardware advancements, including superconducting qubits by IBM  and Google \cite{Kim2023, 48651}, neutral atoms by Pasqal and QuEra \cite{wurtz2023aquila}, and photonic quantum computers by PsiQuantum and Xanadu \cite{Madsen2022}, have created systems surpassing 100 qubits. These systems are reaching the threshold where classical simulation becomes impractical. Consequently, there's a demand for innovative algorithms compatible with the noisy intermediate-scale quantum (NISQ) regime \cite{Preskill_2018}.

Particularly promising are quantum machine learning (QML) algorithms \cite{Cerezo_2022}, which hinge on optimizing variational quantum circuits - a cornerstone of hybrid classical-quantum algorithms \cite{McClean_2016, farhi2014quantum, PhysRevA.92.042303, PhysRevA.94.022309}. These algorithms are noted for their error tolerance and their flexible coherence time and gate requirements, making them suitable for NISQ deployment. However, they confront significant hurdles regarding training, linked to the emergence of barren plateaus in the optimization landscape, which complicates finding the global minimum.

In this paper, we introduce a new {\it hierarchical} variational circuit framework that incrementally learns at different scales. It begins with a small number of qubits and systematically expands to more extensive quantum circuits. Our method not only ensures improved precision and the continuous acquisition of new parameters at a more refined scale but also appears to circumvent the challenge of barren plateaus, which often makes training with 25 or more qubits particularly difficult.

As a case study, we numerically perform distribution loading efficiently using variational circuits. When using a quantum state $\ket{\psi}$ to encode a probability distribution according to its probability amplitudes in the computational basis, the circuit preparing the state is often referred to as a quantum circuit Born machine (QCBM)~\cite{liu2018differentiable,benedetti2019generative, Du:2022kfe, PhysRevResearch.4.043092}. Let $\ket{\psi(\vec{\theta})}$ denote the final state of the variational circuit. In the computational basis, the $\ket{\psi(\vec{\theta})}$ can be written as
\begin{equation}
\ket{\psi(\vec{\theta})} = \sum_{b \in \{0,1\}^{n}} c_b \ket{b}
\end{equation}
where $b$ is summed over bit strings of $n$ qubits. Such a state yields measurement outcomes according to $p(b) = |c_{b}|^{2}$. When using a QCBM to model a distribution, we will map the bitstrings $b$ to the domain of $p(x)$, assumed to be an interval $[x_{0}, x_{f}]$ by interpreting $b$ as a binary fraction. 

For certain distributions, there are analytic results for how to do the loading. For example, normal distributions can be prepared exactly, but their gate count scales exponentially with the number of qubits, making it infeasible for NISQ devices. By using variational circuits and optimization techniques, we can cut down on the required number of gates needed to perform the loading, at the cost of simulation requirements for the optimization. One benefit of this approach would be that once the variational parameters are found, they can be reused in other circuits, serving as a library for distribution loading in other use cases. A typical choice of objective for the variational circuit is the Kullback-Leibler (KL) divergence between the target distribution and the generated distribution $q_\theta$.
\begin{equation}
  \label{eq:kl}
  KL(p|q_{\theta}) = \sum_{x} p(x) \log \left(\frac{p(x)}{q_{\theta}(x)}\right) \, .
\end{equation}

% Our second aim with distribution loading is then to perform manipulations, such as multiplication or convolutions of such distributions, for the purposes of efficiently loading more complex distributions which are out of reach for classical methods. Similar to the naive application of distribution loading, finding variational circuits whose depth is shallow will be critical for finding near term applications of these methods given current error rates.

% \HG{To do for Vince: Add discussion on Barean Plateaus and claims of non-trainability of variational methods. Discuss using non-observable loss function as a tools for quantum data loading. Provide relevant references.}

The primary objective is to train QCBM models by minimizing the KL divergence. Traditional finite difference methods, hindered by a quadratic time complexity with respect to the number of circuit parameters, were inefficient for large-scale applications. To tackle this, we implemented an adjoint derivative technique \cite{jones2020efficient}, analogous to backpropagation in neural networks, but tailored for variational quantum circuits. This approach significantly streamlines the process with its linear time complexity in circuit parameters, greatly enhancing the scalability and efficiency of QCBM training.

Given the state of noisy intermediate-scale hardware, lots of attention has been focused on a hybrid classical-quantum setup where a variational quantum circuit is used with some classical processing to handle gradient updates. Often, the quantity being minimized is the expectation value of an observable, such as the energy of a Hamiltonian. It was shown that gradients of such expectation values in random circuits (or equivalently sufficiently deep variational circuits) have a concentration effect around zero, often referred to as barren plateaus~\cite{McClean_2018}. This effect also had an exponential scaling as a function of the number of qubits. Combined, both paint a pessimistic picture about the feasibility of effectively optimizing variational circuits with random initializations, in contrast to the classical case of neural networks. Further works~\cite{Cerezo_2021,Holmes2022Expressibility,Patti2021entanglementbp,Wang:2020yjh} attribute concentration effects to the locality of the observables, expressibility of the circuit, the entanglement of the state, and the presence of noise. 

Motivated by quantum devices with larger qubit counts, we present novel results for learning to load a 3-dimensional multivariate Gaussian distribution on 27 qubits, the highest numerical experiment for such a task to our knowledge. A priori, larger qubit numbers yield unknown challenges when it comes to expressibility, whether a given variational circuit has the flexibility to encode the distribution of interest, and also the learning, where deeper circuits are known to exhibit a concentration of gradients. Since our loss is not based on the expectation value of an observable, the barren plateau arguments do not strictly apply. Barren plateau results for QCBMs are generally lacking, though see~\cite{Rudolph:2023iqf} for recent progress in certain settings. Our learning results do not contradict this as we assume access to the underlying state with numerics. 

% Optional: talk about not needing to scale this approach since many more digits are not necessary
% We can consider this as a subroutine for interesting datasets and then do further quantum manipulations on them
Lastly, we have developed hardware-specific variational ansatzes and employed a hierarchical learning approach for training hardware-specific circuits on classical computers, which were then executed on IBM Quantum Computers \cite{Mooney_2021}. This was done to assess the performance of QCBM model. We utilized IBM's 7-qubit Lagos and 27-qubit Algiers machines, operating circuits in three distinct modes: vanilla IBM (with default Sampler settings), Fire Opal-Optimized IBM (utilizing Q-CTRL's Fire Opal software \cite{PhysRevApplied.20.024034}), and an ideal classical simulator. Our primary objective was to compare the Total Variation (TV) distance, an indicator of approximation accuracy, across these modes.

During the 7-qubit experiments on IBM Lagos, we ran four circuits, each designed to prepare the same normal distribution, albeit with different numbers of qubits and two-qubit gates. The outcomes indicated a marked improvement in approximation accuracy when using the Fire Opal optimization. In the 27-qubit experiments conducted on IBM Algiers, we tested four circuits aimed at generating a bimodal mixed Gaussian distribution. Here, the enhancements brought about by Fire Opal optimization were particularly significant in circuits that included more than 22 two-qubit gates.

This study highlights the effectiveness of using advanced GPU simulators and hierarchical learning for circuit training. Significantly, it proves the possibility of accurately deploying these circuits on actual quantum hardware with the use of powerful control tools like Fire Opal. While QCBMs may not achieve exponential speedup due to the necessity of full state tomography, the approach to data loading presented here could be valuable as a component in larger quantum circuit design. This approach facilitates the efficient transfer of segmented datasets into the quantum device, setting the stage for subsequent operations and manipulations with quantum circuit methods. 

The potential of hierarchical learning extends beyond QCBMs. We anticipate that this technique could enhance the performance of various variational circuits, offering a means to achieve superior approximation in a wider range of quantum computing applications.

\section{Methods}
We review two main accelerants that enable the training of large-scale QCBMs. The first is an analytical insight into how QCBM measurements map to samples. The basic idea is to train the most significant qubits of a QCBM first. The second is a computational advance that allows for efficient simulation of gradients. We extend the well-known adjoint technique to non-observable loss functions and implement it in our simulations.

\subsection{Hierarchical Learning}

\begin{figure}
  \centering
  \includegraphics[width=.9\linewidth]{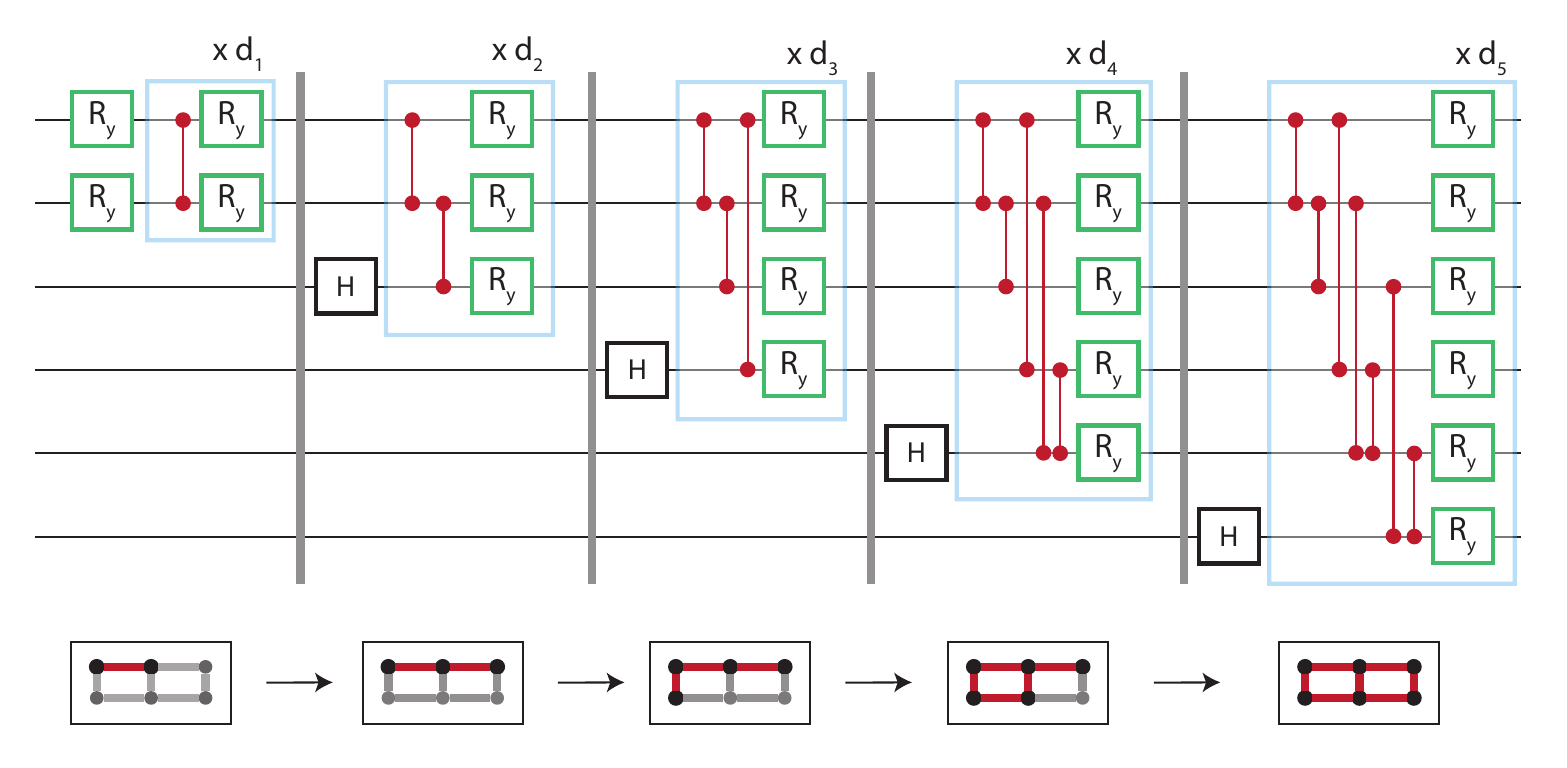}
  \caption{A circuit ansatz with hierarchical architecture on 6 qubits laid out on a grid. Vertical gray lines on the circuit indicate a partial circuit which is variationally learned before using it as input to the next sub-circuit. Starting with two active qubits, we learn variational parameters that approximate the distribution sampled at $2^{2}=4$ data points. Red two-qubit gates correspond to the variational RZZ gates. After training on the two qubit VQC, learned parameters are used as the starting parameters for a 3 qubit VQC with the third qubit initialized in the $\ket{+}$ state and all new parameters are initialized to 0 rather than randomly. The bottom row visualizes the active qubits (black), inactive qubits (gray), and active connectivity (red) for each sub-circuit of the hierarchical VQC. }\label{fig:hierarchical}
\end{figure}

Due to the difficulty in training deep circuits, we propose a novel method for QCBMs with a large number of qubits. In this section, we explain the structure behind our hierarchical approach, visualized in Fig.~\ref{fig:hierarchical}. As the number of qubits grows, variational circuits require more gates to capture the richer pattern of correlation that can emerge. The primary insight is to exploit the structure of the mapping between bitstring measurements and the samples they represent. In particular, for smooth distributions, the correlations between the most significant (qu)bits play an outsized role. Thus, we propose doing the QCBM training in stages, starting with a relatively small number of qubits. The variational parameters of the QCBM are then trained on a coarse-grained version of the distribution. This then seeds a larger variational circuit with more qubits and gates. New parameters are initialized to be 0 rather than randomly instantiated. In this work, we make use of the standard hardware efficient ansatz, but the hierarchical strategy can be combined with other variational ansatzes. We refer to qubits that have participated in the training to be \textit{active} and \textit{latent} otherwise.

\begin{figure}
  \centering
  \includegraphics[width=.8\linewidth]{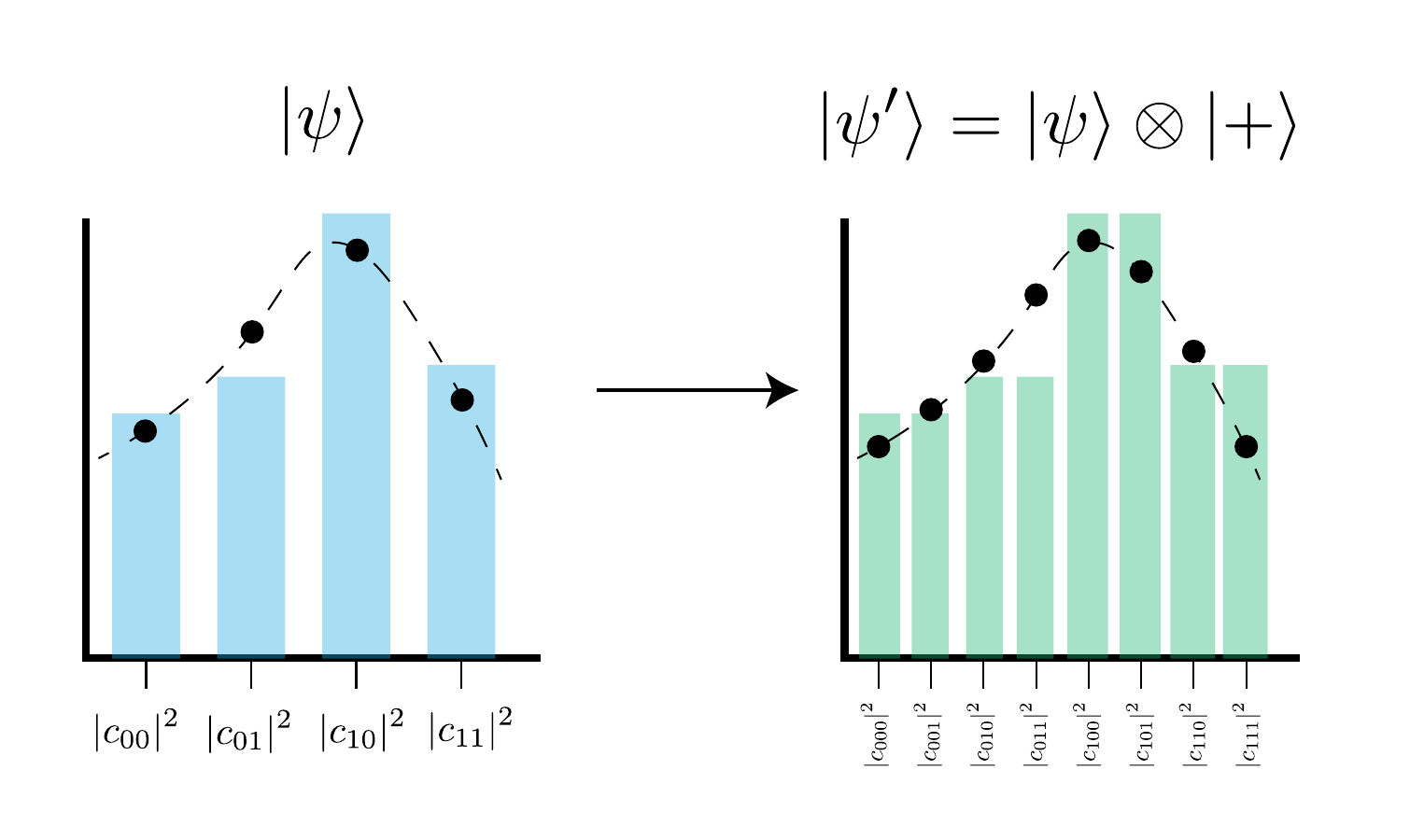}
  \caption{Adding a qubit initialized in the $\ket{+}$ state to a QCBM yields a distribution that is piece-wise constant. When the last qubit is interpreted as the least significant bit, the probability mass gets split into neighboring bins. The underlying distribution (black), which was sampled at discrete points must also be renormalized so that the total sums to 1.}\label{fig:expansion}
\end{figure}

Qubits added to the circuit at later stages are initialized in the $\ket{+} = \frac{\ket{0} + \ket{1}}{\sqrt{2}}$ state. When interpreted as the least significant bit, this leads to an equal amplitude for bitstrings that share the same prefix. Thus, adding a qubit in the $\ket{+}$ state to an $n$ qubit QCBM yields a QCBM whose mass distribution is split into neighboring bins. This will generally give a good approximation to the more fine-grained distribution if the underlying distribution is smooth. See Fig.~\ref{fig:expansion}.

To aid the interpretability of the QCBM performance, we also track the total variational (TV) distance between the target distribution $p$ and the learned distribution $q_{\theta}$.

\begin{equation}
  \label{eq:tvd}
  \TV(p,q_{\theta}) = \frac{1}{2} \sum_{x} |p(x) -q_{\theta}(x)| \, .
\end{equation}

Because we will compare probability mass distributions with different support, arising when comparing QCBMs of different qubit numbers, we introduce the following notation. Let $\TV_{n}(p, q_{\theta})$ denote the TV distance where both distributions are sampled on $2^n$ points. For example, this naturally occurs if $q_{\theta}$ is a QCBM on $n$ qubits. If the target distribution has support on $2^{m}$ points with $m > n$, then $\TV_{m}(p,q_\theta)$ would be computed with the distribution arising from the addition of $m-n$ qubits in the $\ket{+}$ state. In the case where $m < n$, the distribution is coarse grained by marginalizing over the least significant qubits. This amounts to adding the probability mass of neighboring bins $n-m$ times.

This is particularly important when comparing the performance at different stages of the training. As an extreme example, if one uses a QCBM of a single qubit, the naive TV$_1$ distance would be a sum of two terms. A single RY gate would suffice to drive $\TV_{1}$ to 0, yet we would not say that this fully captures the underlying distribution. Thus in our larger experiments, we will track the TV distance with a resolution of the largest qubit size, e.g. $\TV_{n}$ regardless of how many active qubits there are. We show an illustrative set of data that compares the $\TV$ distance at varying resolutions for a learned distribution in App.~\ref{app:tvd}.

% The addition of the least significant bit forms the basis for our novel circuit design we designate as hierarchical learning. Because the most significant bits play an outsized-role, it makes sense to first learn a coarse grained version of the distribution as a starting point. One can thus repeat this procedure of training a QCBM on $n$ qubits before expanding the next layer of the variational ansatz to include $n+1$ qubits until the desired resolution is achieved. This ansatz is both more resource efficient and, as we showcase in later sections, dramatically increases performance when it comes to larger qubit numbers. See Figure~\ref{fig:hierarchical} for a visualization of the circuit.

\subsection{Adjoint Derivative}\label{ssec:adjoint}
The training of the QCBM is performed by minimizing the KL divergence between the final produced distribution, a function of the wavefunction coefficients, and the target distribution. To compute gradient elements, we want to know how the loss varies with respect to each of the $M$ variational gate parameters. Since gate evolution produces unitary dynamics, the effect on the state vector for the circuit is repeated matrix multiplication. Thus, one can apply similar techniques of backpropagation to the circuit as one would do for a neural network. However, a naive application of the chain rule leads to a memory requirement that scales with the size of the matrices involved, which in our case will be $O(2^{2n})$.

In the case of gradients of the expectation value of an observable $\langle B \rangle_{\psi(\vec{\theta})}$, it was shown in \cite{jones2020efficient,Luo2020yaojlextensible} that one can efficiently compute gradient elements for $M$ parameters in $O(2^{n})$ space and $O(2^{n} M)$ time as opposed to $O(2^{n} M^{2})$ time required for finite differences. Assuming that each gate is easily differentiable, as is the case for gates that are exponentials of Pauli operators, $\partial_{\theta_{i}}$ acts on the bra $\bra{\psi(\vec{\theta})}$ and the ket $\ket{\psi(\vec{\theta})}$ and they appear as complex conjugates of each other. Another key insight is that each gradient element itself can be cast as the inner product of two state vectors which involve some subset of unitaries applied to $\ket{\psi_{0}}$ and $\bra{\psi_{0}}$. These state vectors are successively updated for each gate parameter but do not require intermediate storage of any of the unitaries in the circuit. \footnote{See Pennylane tutorial for a nice explanation.} This is referred to as the adjoint method for computing gradients since the updates to $\ket{\psi}$ and $\bra{\psi}$ involve applications of $U_{i}$ and $U_{i}^{\dag}$.

As alluded to previously, however, arbitrary functions of the amplitudes are not observable expectation values in a state. Nevertheless, this approach can be leveraged to compute numerical gradients efficiently if one can recast the gradient computation as an inner product between two state vectors. We proceed to show how this can be done in the case of the L2 loss of the QCBM. Let $L(\vec{\theta}) $ be the loss

\begin{equation}
  \label{eq:1}
  L(\vec{\theta}) = \sum_{x} (q_{\vec{\theta}}(x) - p(x))^{2}
\end{equation}

Taking the derivative with respect to the $i$-th parameter, we get

\begin{equation}
  \label{eq:2}
  \frac{\partial L(\vec{\theta})}{\partial \theta_{i}} = \sum_{x} (q_{\vec{\theta}}(x) - p(x))\frac{\partial q(x)}{\partial \theta_i}
\end{equation}

Since $q_{\theta}(x) = |c_{x}|^{2 } = c_{x} \bar{c_{x}}$,
\begin{equation}
  \frac{\partial q(x)}{\partial \theta_i} = \frac{\partial c_{x}}{\partial \theta_i} \bar{c_{x}} + \frac{\partial \bar{c_{x}}}{\partial \theta_i} c_{x} = 2 \Re\left[ \frac{\partial c_{x}}{\partial \theta_i} \bar{c_{x}}\right] \, .
\end{equation}

Let us focus on the term $\frac{\partial c_{x}}{\partial \theta_i} $, where $c_{x}$ are precisely the coefficients of the wavefunction

\begin{align}
  \label{eq:3}
  \frac{\partial}{\partial \theta_i}  \ket{\psi_{n}} &=   \frac{\partial}{\partial \theta_i}  U_{n}U_{n-1} \ldots U_{1}\ket{\psi_{0}} \\
                              &= U_{n} \ldots U_{i+1}U'_{i}U_{i-1} \ldots U_{1} \ket{\psi_{0}} \\
                              &= U_{n} \ldots U_{i+1}(iG U_{i})U_{i-1}\ldots U_{1} \ket{\psi_{0}} \, ,
\end{align}

where we have assumed that each $U_{i}$ is generated by some operator $G$ as in $U_{i} = \exp(iG\theta)$.
Thus, our gradient element $\frac{\partial L}{\partial \theta_{i}}$ can be written as an inner product

\begin{align}
  \frac{\partial L(\vec{\theta})}{\partial \theta_{i}} &= \sum_{x} 2 \Re \left[\left((q_{\vec{\theta}}(x) - p(x))\bar{c_{x}}\right) \left(\frac{\partial c_x}{\partial \theta_i}\right) \right] \\
  &=  2 \Re[\bra{b_{n}} U_{n} \ldots U_{i+1} (iGU_{i}) U_{i-1} \ldots U_{1}\ket{\psi_{0}}] \\
  &=  2 \Re[\bra{U^{\dag}_{i+1} \ldots U_{n}^{\dag}b_{n}}    iG U_{i} U_{i-1} \ldots U_{1}\ket{\psi_{0}} ]\\
  &=  2 \Re[\bra{b_{i+1}}    iG \ket{\psi_{i}} ]
\end{align}

with $\ket{b_{n}} = \diag(q_{\theta}(x)-p(x))\ket{\psi_{n}}$.

For a different choice of loss that depends on all the coefficients, one simply needs to change the initial ket $\ket{b_{n}}$. In the case of the KL divergence, $\ket{b_{n}} =\diag(\log\frac{q_{\theta}(x)}{p(x)} + p(x))\ket{\psi_{n}}$.

The main upshot to this technique is the scaling with the number of parameterized gates, $M$. Using a method of finite differences, one needs $O(M^{2})$ gate applications. Each circuit computation requires $O(M)$ gates simply to evaluate, and each of the $M$ parameters needs a full circuit evaluation. With the adjoint method, $\psi_{n}$ is calculated in $O(M)$ time, and each of the gradient elements can be obtained by successively applying $U^{\dag}_{i}$ to update $\ket{\psi}$ and $\ket{b}$. Thus the total number of gate applications is linear, once on the forward pass and once on the backward pass.

% \subsection{Hybrid GPU Acceleration}
% % To perform our numerics, we utilize GPU accelerated numerics to perform parallelized
% maybe ask Hayk to fill in here

% \begin{figure}
%   \centering

%   \includegraphics[width=.8\textwidth]{figures/hea-vqc}
%   \caption{An example HEA we use to load distributions. Alternating layers of single qubit rotations followed by two qubit entanglers are repeated $d$ times. The connectivity of qubits will determine the topology of the entangling gates.}
%   \label{fig:hea-vqc}
% \end{figure}

% One option to circumvent these problems with the KL divergence is to use the \textit{kernel trick}, where the distance between the output and target distribution is calculated in some auxiliary feature space rather than the usual trace norm. This approach was discussed and used by~\cite{liu2018differentiable}. The choice of such a feature space, or equivalently, the kernel, is an interesting avenue for future investigation.

\section{Distribution Loading}\label{sec:results}
We now turn to the application of QCBMs to load distributions. We begin with the task of loading a univariate distribution to investigate how the circuit connectivity affects QCBM performance. Importantly, we find that a hardware-efficient ansatz (HEA)~\cite{kandala2017hardware} on the grid performs better than a similar circuit with ring connectivity. Equipped with this knowledge, we build and train a hierarchical circuit to load a 3-dimensional multivariate Gaussian distribution on 27 qubits.

\subsection{The Role of Connectivity for Small Scale QCBMs}
\begin{figure}

  \centering
  \includegraphics[width=.442\linewidth]{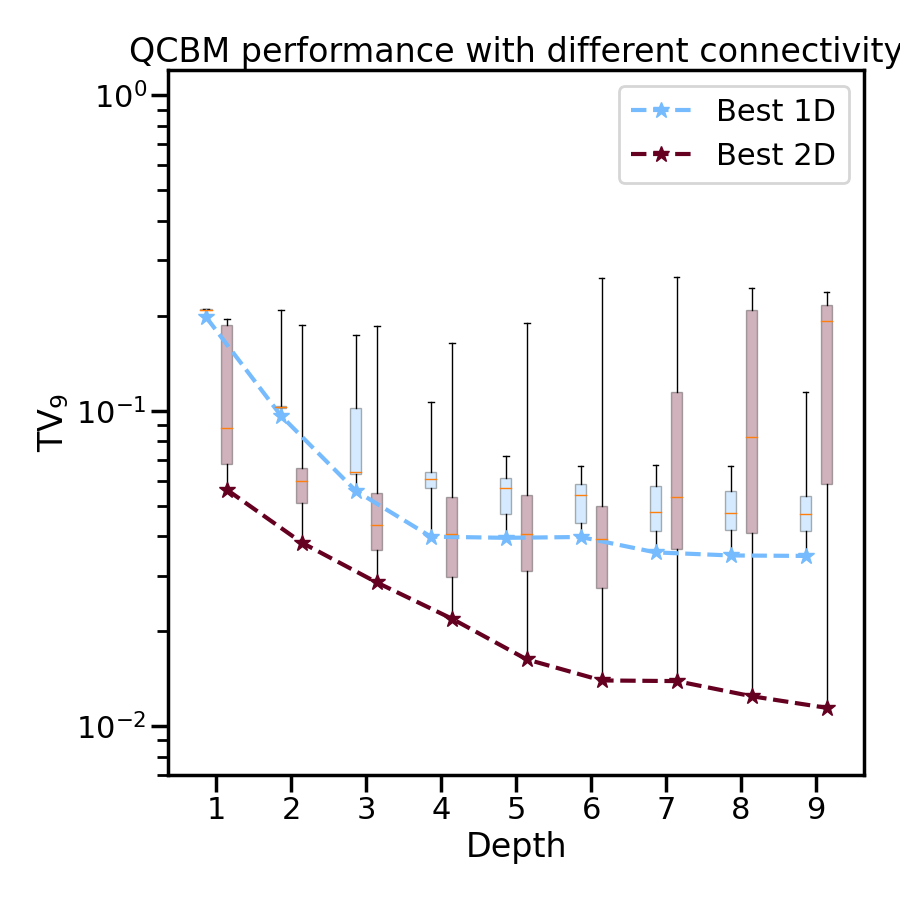}
  \includegraphics[width=.49\linewidth]{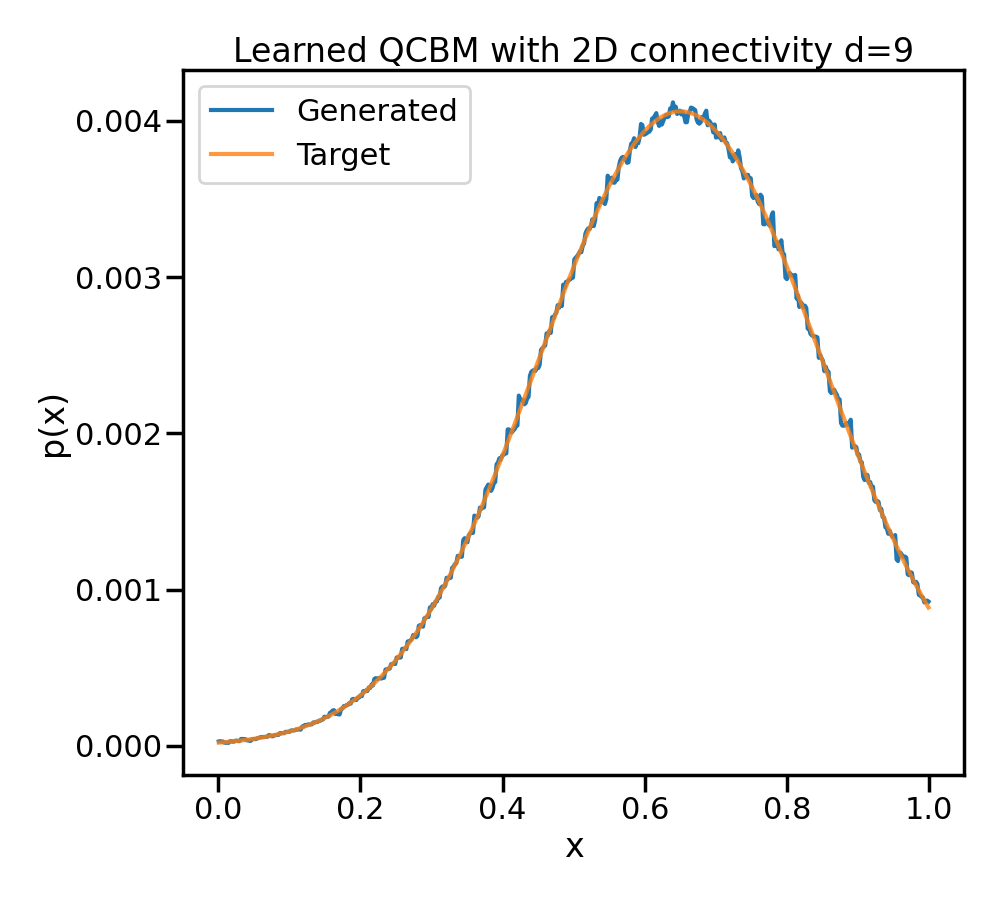}

  \caption{a) Comparing the performance of different connectivity for QCBMs on 9 qubits as a function of depth. For a given number of layers, the learning procedure was repeated with 50 different seeds for the initial variational parameters. Colored rectangles denote the 25-75th percentile ranges with the whiskers extending to minimum and maximum. Orange lines denote the median. For 1D connectivity, additional layers very quickly lead to small marginal improvements.  b) The best-resulting distribution with 2D topology. Here $n=9$ qubits were used with a depth $d=9$ circuit resulting in a TV$_9$ of $\sim 0.5 \%$.}\label{fig:1d_gaussian_fit}
\end{figure}

To illustrate the role of variational circuit design choices, we start with a simplified setup of our large-scale experiments. Consider a univariate Gaussian on the interval $[0, 1]$ with mean and variance $\mu=0.65, \sigma^{2} =0.04$. Motivated by existing hardware connectivity, we compare the performance of two architectures of variational circuits, 1D connectivity as in a ring, and 2D connectivity on a square grid. Here we fix the number of qubits at $9$, resulting in a discrete approximation over $2^9=512$ points.

In Fig.~\ref{fig:1d_gaussian_fit}, we report the TV distance of QCBMs as the number of layers grows. Each layer consists of single qubit $RY$ gates on all qubits and $RZZ$ between connected qubits. For a given circuit, we run 1000 epochs of gradient descent with the Adam optimizer~\cite{kingma2017adam}. Because there is variation depending on the random initialization of parameters, we repeat each experiment 50 times and calculate both the minimum TV distance as well as the inter-quartile range to show the variation in performance.

We find there is a rather dramatic effect of connectivity on QCBM performance. Shallow 1D circuits are not expressive enough, and adding more layers beyond 4 does not lead to a better TV distance. By contrast, the 2D circuits do not seem to plateau. As the depth increases, one sees a sensitive dependence on the initial parameters as denoted by the variance of TV distance over multiple initializations.

\subsection{Large Scale Multivariate Distribution Loading}

\begin{figure}
  \centering
  \includegraphics[width=.7\linewidth]{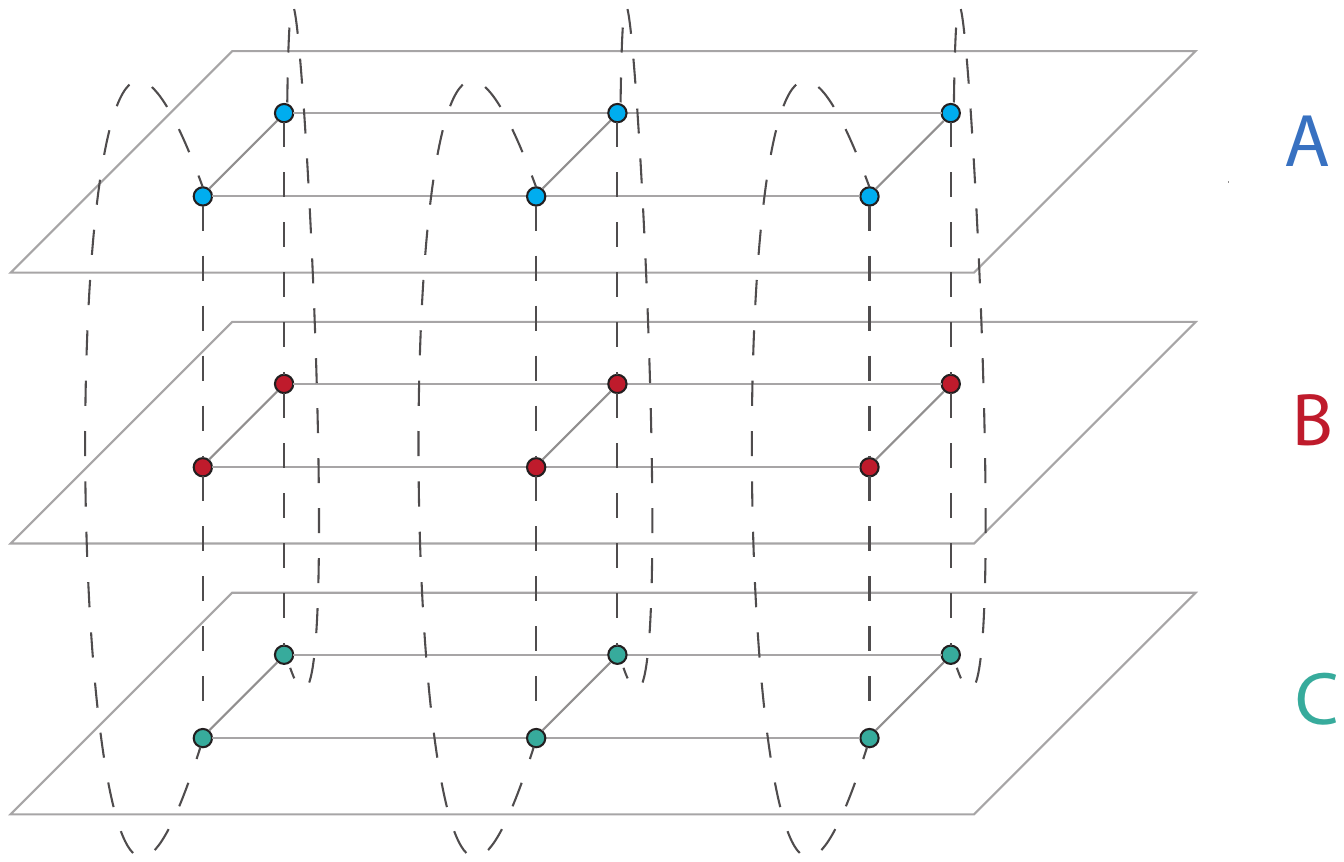}
  \caption{Overview of connectivity between 3 sets of qubits $A,B,C$ that represent the domains of the 3 variables of interest. Within each set, they are assumed to be on a 2D grid. Corresponding qubits between each set are also connected. Here a line between two qubits indicates that the variational circuit includes a two-qubit gate acting between them.}\label{fig:2d_connectivity}
\end{figure}

In this section, we share the techniques that enable us to efficiently learn variational parameters for a multivariate Gaussian distribution on 27 qubits. Computationally, this was made feasible by the use of the adjoint method for computing gradients. We extend this technique to the case of non-observable loss functions. We employ a hardware efficient ansatz and show that the hierarchical learning strategy outperforms learning all of the variational parameters at the same time for the same final circuit.

The distribution we load is a 3D multivariate Gaussian with the following mean and covariance parameters with support restricted to the unit cube $[0, 1]^{3}$.

\begin{align}\label{eq:multi_params}
\vec{\mu} &= \begin{bmatrix}
0.5 & 0.3 & 0.7
\end{bmatrix}    \\
\Sigma &= \begin{bmatrix}
0.2 & -.1 & -.1\\
-.1 & 0.1 & 0 \\
-.1 & 0 & .3
\end{bmatrix}
\end{align}

The unnormalized\footnote{The normalization constant will be different since we are restricting the support.} PDF is then given by
\begin{equation}
    p(\vec{x}) = \frac{1}{\sqrt{(2\pi)^k \text{det}(\Sigma)}} \exp\left[-\frac{1}{2} (\vec{x}-\vec{\mu})^T \Sigma^{-1} (\vec{x}-\vec{\mu}) \right]
\end{equation}

Let us discuss how the QCBM mapping works for the multivariate case as well as our circuit architecture. Let $A_i$ denote the $i$-th qubit that maps to the domain of $X$. Analogously to the univariate case, we separate the length $n$ bitstring into 3 equal size partitions, $\ket{b} = \ket{b_A b_B b_C}$. Each of the sets of qubits is presumed to lie on a grid with nearest-neighbor interactions. The form of our variational circuit will be a HEA with this connectivity as well as interactions between sets, e.g. connecting $A_i$ and $B_i$. See Fig.~\ref{fig:2d_connectivity}. For the hierarchical structure, qubits $A_1, B_1,$ and $C_1$ are the most significant bits. We will add qubits in multiples of 3 so the number of active qubits in each set is the same.

\begin{figure} [H]
  \centering
  \includegraphics[width=.55\linewidth]{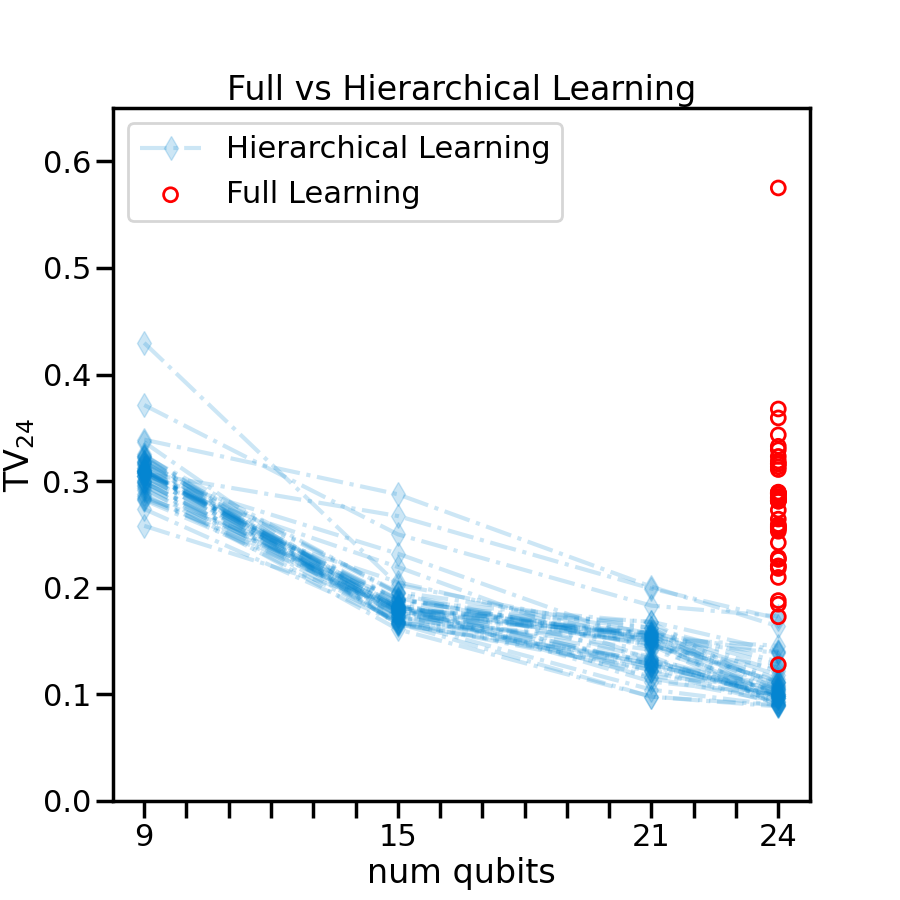}
  \caption{Learning hierarchically performs better both on average and in absolute terms versus learning all the parameters of the final hierarchical circuit. Blue dashed lines indicate performance as more layers are added to the hierarchical ansatz. Red dots indicate performance when the learning is performed on the same final variational circuit with fully random parameters. Here we learn the same three-dimensional multivariate distribution as in Eq.~\ref{eq:multi_params}. For the hierarchical ansatz, we start with three qubits per variable and random parameter initialization. An additional two qubits are added with new variational parameters initialized to 0. This is repeated until the circuit expands to 24 qubits. }\label{fig:layerized_parameters}
\end{figure}

To first highlight the advantage of the hierarchical approach, we perform the QCBM training in two ways. The first utilizes the hierarchical approach where learning takes place in four stages, starting with 9 active qubits and expanding up to 24. The second (non-hierarchical) approach takes the same final circuit ansatz on 24 qubits and learns all parameters from scratch. In the hierarchical case, we report $TV_{24}$ for different numbers of active qubits. Repeating this procedure for various seeds, we see that the hierarchical strategy yields the best QCBM overall and on average. See Fig.~\ref{fig:layerized_parameters}.

\begin{figure}
  \centering
  \includegraphics[width=.55\linewidth]{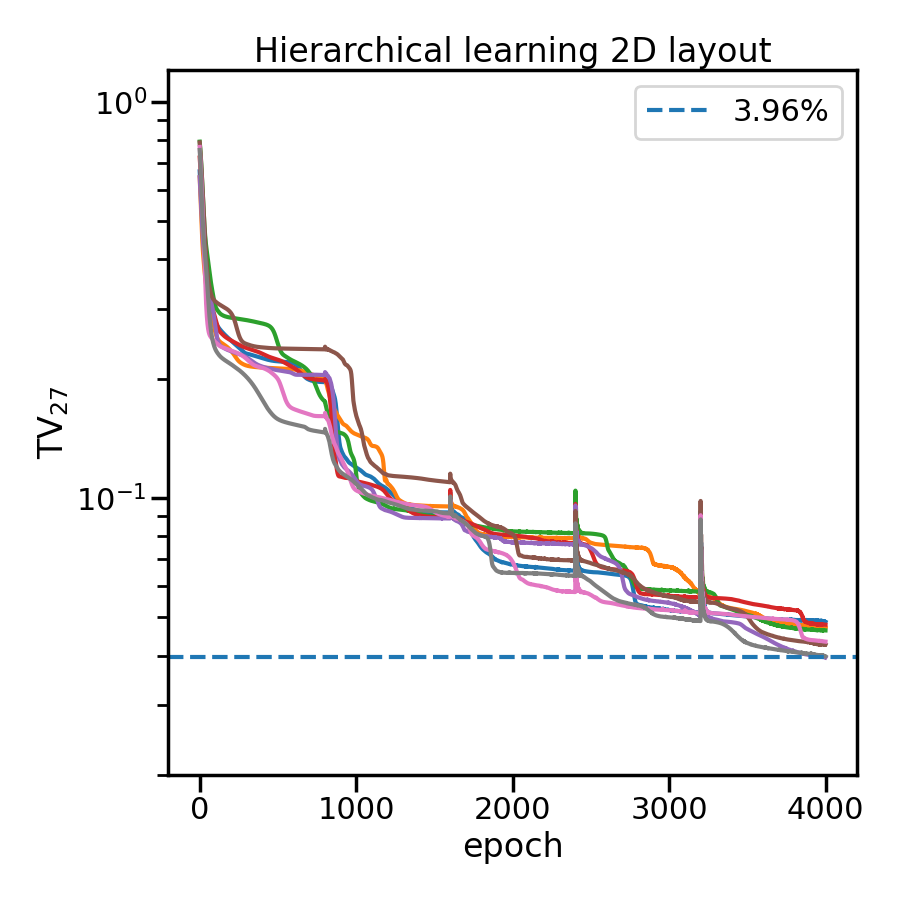}
  \caption{TV$_{27}$ distance when training a hierarchical QCBM to reproduce a 3D multivariate gaussian distribution using 27 qubits. Each of the variables is assigned a set of nine qubits arranged on a 3x3 grid. The connectivity of the HEA is similar to that of Fig.~\ref{fig:2d_connectivity}.
  }\label{fig:3d-learning}
\end{figure}
\begin{figure}
  \centering
  \includegraphics[width=0.8\linewidth]{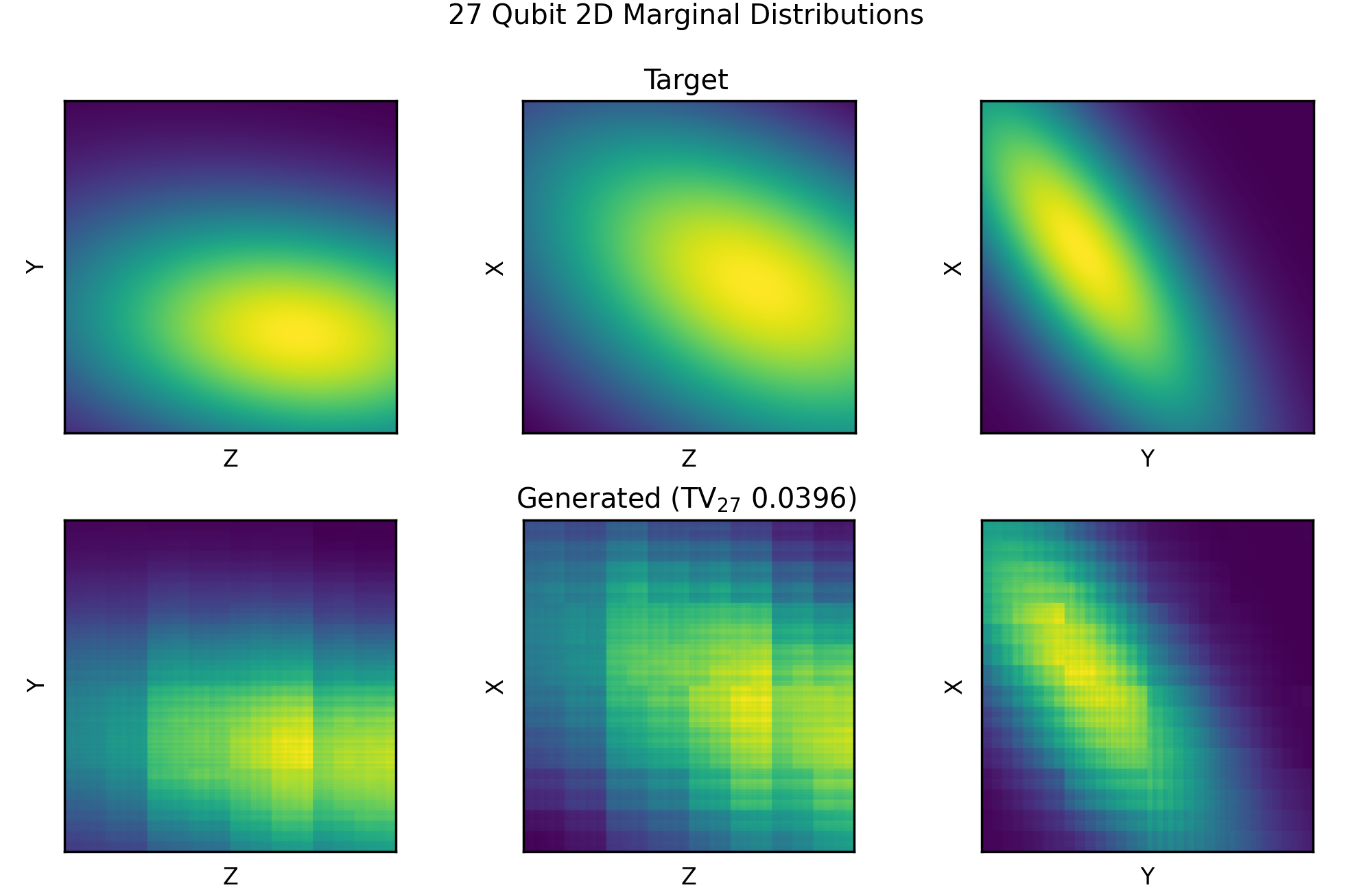}
  \caption{Comparing marginal distributions of two variables from the target distribution (top) to the best-learned distribution (bottom). }\label{fig:3d_marginals}
\end{figure}

Finally, we train hierarchical circuits to load the multivariate Gaussian on 27 qubits. The learning takes place in 5 stages, starting from 9 qubits. We track the $\TV_{27}$ in Fig.~\ref{fig:3d-learning}. Notably, as in Fig.~\ref{fig:layerized_parameters}, the variance in performance is relatively small, implying that this strategy is useful when the simulation is limited and many seeds cannot be run. We show a visualization of the learned distribution by projecting onto marginal distributions of two variables in Fig.~\ref{fig:3d_marginals}.

% Figure~\ref{fig:3d-learning} shows the result of QCBM learning of the multivariate Gaussian using all-to-all connectivity in the variational circuit. The total number of variational gates $M$ for 24-27 qubits and $d=1$ layers of all-to-all gates is several hundred parameters. Because each gradient calculation requires $O(M)$ calculations, we make use of parallelized GPU calculations. For 24 qubits, we see quite a difference in the final L1 loss as a result of different seeds. More layers tended to yield slightly better approximations, but this was subdominant to having a good seed.

% The other approach we took was to have a layerized approach that featured a lower degree of connectivity. Let $A, B, C$ denote the set of qubits responsible for describing bits of the variables $x, y, z$. Qubits within $A$, $A_{1}, \ldots, A_{n/3}$ were treated as living on a 2D grid. The variational ansatz then included $R_{ZZ}$ gates between neighbors. Between $A,B$ and $C$, we connected corresponding bits (e.g. $A_{i}$ with $B_{i}$) as in Figure~\ref{fig:2d_connectivity}.

\subsection{Classical Hardware Used in Simulation}

We have been using Nvidia V100 and A100 GPUs for the experiments presented in this paper. For training the QML circuits and optimizing the parameters we have been using Pennylane's QML modules~\cite{Bergholm:2018cyq}. As mentioned in section \ref{ssec:adjoint} we achieved a significant speedup over naive QML differentiation methods by implementing our adjoint gradient method for KL divergence loss. 
Instead of scaling quadratically with the number of parameters $M$, our approach scales linearly - resulting in bigger speedups for larger $M$. Having said that, single circuit evaluation with adjoint technique is 10x slower compared to a vanilla circuit evaluation, so all together we get $M/10$ overall speedup.
Specifically, on Nvidia A100 GPU using Pennylane, training a 30 qubit, 1000 parameter circuit for 150 epochs would take $\sim 730$ hours with a naive method, and only $\sim 7.3$ hours using our adjoint trick.
This means that large intractable experiments that would take a month become possible to carry out overnight.

\section{IBM deployment of QCBM models} \label{sec:IBM-Fire}
In this section, we have used the hierarchical learning method to train quantum circuits on classical devices and executed those circuits on IBM hardware in different modes to see its performance on modern Quantum Computers. 
Specifically, we compare the TV distance across 3 different modes - vanilla IBM, Fire Opal-Optimized IBM, and ideal classical simulator.
Here vanilla IBM means executing the circuit using IBM Sampler with default settings i.e. \textit{resilience\_level}=1 and \textit{optimization\_level}=3. Fire Opal Optimized means running the circuit through Q-CTRL's proprietary Fire Opal software~\cite{PhysRevApplied.20.024034}. The Ideal Simulator shows how good our circuits are approximating the desired distribution have we had an ideal quantum computer. 
We carried out this experiment on 7 qubit IBM Lagos and 27 qubit IBM Algiers for multiple circuit designs.
We used the hardware connectivity in our ansatz to come up with optimal circuits requiring minimal number of 2-qubit gates. 

\subsection{7 qubit Experiments: IBM Lagos}
\begin{figure}[t]
  \centering
  \includegraphics[width=0.8\linewidth]{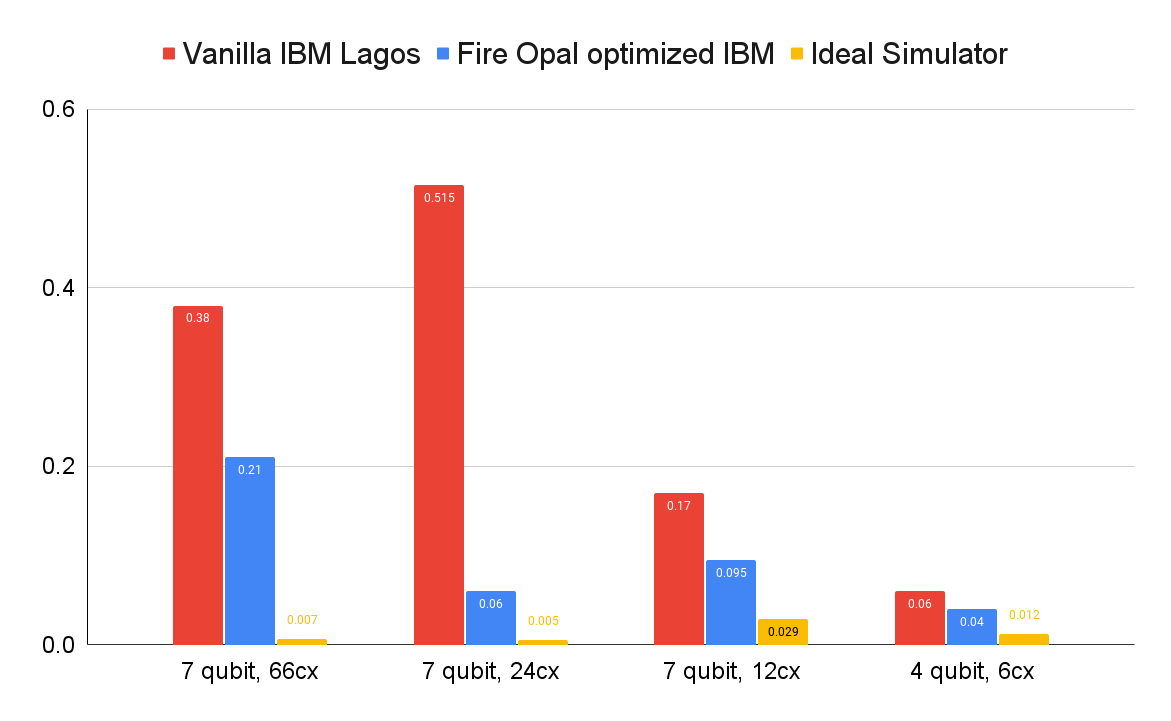}
  \caption{$TV_4$ for 4 different circuits in 3 different device modes. We can see how Fire Opal optimized runs have $1.5-8$x advantage over vanilla IBM.}\label{fig:firopal_vs_ibm_lagos}
\end{figure}

On 7-qubit IBM Lagos we ran 4 circuits - all preparing the same normal distribution - with different number of qubits and 2-qubit gates. The only type of 2-qubit gates used in these circuits were CX gates and their number ranges from 6 to 66. Since the number of qubits is low here we do not apply hierarchical learning and instead train the circuit as a whole.  To compare approximation accuracy across different qubit-numbered circuits we will use the $TV_{4}$ metric.
Both for IBM and Fire Opal runs we repeated the experiment 5 times and took the best outcome.
The results are in Fig.~\ref{fig:firopal_vs_ibm_lagos}. As we can see there is a huge improvement when using Fire Opal optimized circuits.

\begin{figure}[H]
  \centering
  \includegraphics[width=1.0\linewidth]{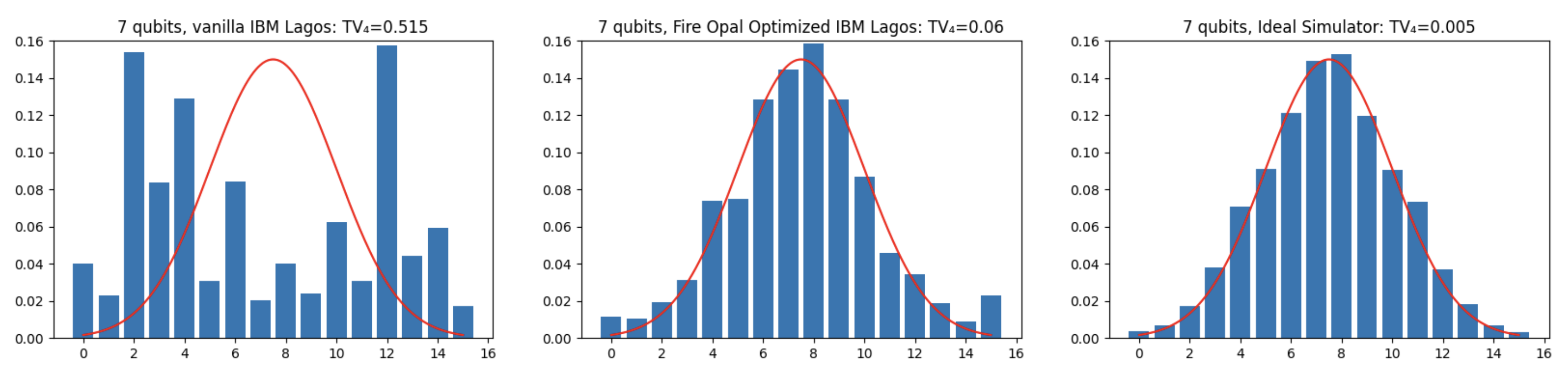}
  \caption{3 visualizations of the same 7 qubit 24CX circuit. The red line is the underlying true distribution. We see how the accuracy of the visualization get better from left to right, as $TV_4$ decreases. The quantum device used here is the 7 qubit IBM Lagos}\label{fig:firopal_vs_ibm_lagos_vis}
\end{figure}

To illustrate the quality of approximations for these 3 different modes we can look at figure \ref{fig:firopal_vs_ibm_lagos_vis} that shows the approximated distribution stored in bitstring amplitudes. This figure is from the 7-qubit, 24CX circuit. We are visualizing only the 4 most significant qubits - hence $16$ indices. We can visually see how the approximation gets better from vanilla IBM to Fire Opal-Optimized to the ideal simulator.

\subsection{27 qubit Experiments: IBM Algiers}
\begin{figure}
  \centering
  \includegraphics[width=0.9\linewidth]{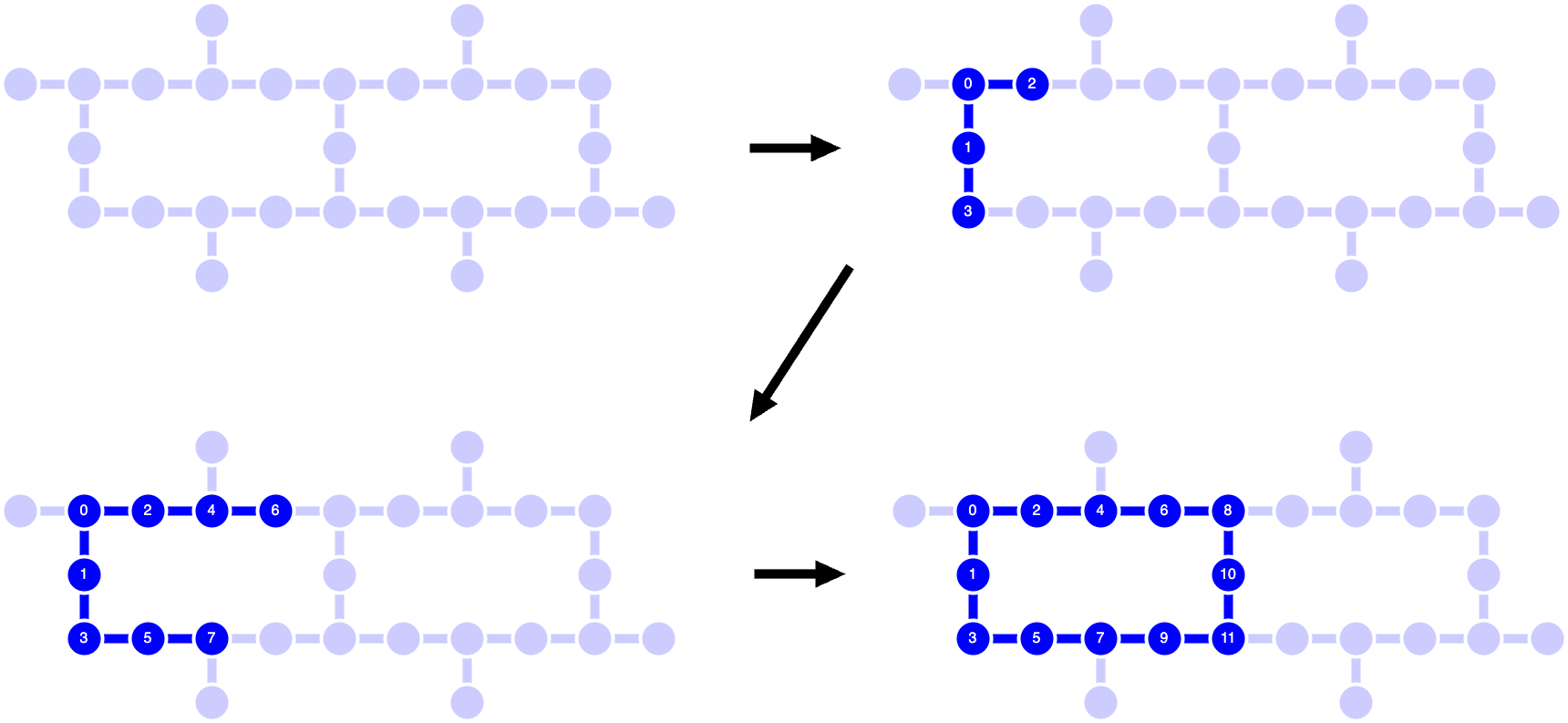}
  \caption{The diagram for the hierarchical ansatz used in training of 12 qubit, 22cx circuit. The underlying diagram is IBM Algiers connectivity map.}\label{fig:ibm_algiers_diagram}
\end{figure}

On 27 qubit IBM Algiers we run 4 circuits - all preparing the same bimodal mixed Gaussian distribution - with different number of qubits and CX gates.
We apply hierarchical learning here by starting with 4 qubits and growing the circuit by another 4 on each iteration. Figure \ref{fig:ibm_algiers_diagram} shows the iterative process used for the hierarchical learning of the 12 qubit circuit. The other 3 circuits have similar diagrams, e.g. the ansatz is chosen to match the connectivity of IBM Algiers.

\begin{figure}
  \centering
  \includegraphics[width=0.8\linewidth]{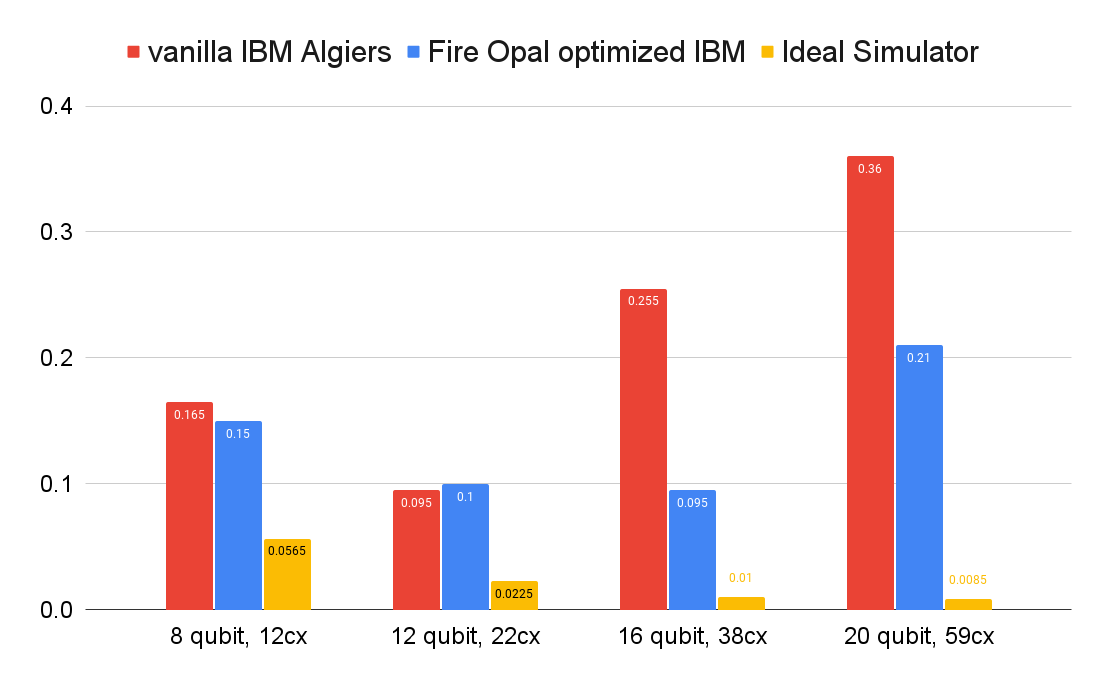}
  \caption{$TV_5$ for 4 different circuits in 3 different device modes. We can see how Fire Opal optimized runs perform much better than vanilla IBM runs for larger circuits}\label{fig:firopal_vs_ibm_algiers}
\end{figure}

Since we are dealing with a bimodal distribution here with two peaks, we decided to use twice as many buckets for visualization and measuring accuracy, hence we are using $TV_5$ distance here instead of $TV_4$.
The results are in figure \ref{fig:firopal_vs_ibm_algiers}. We can see how this time Fire Opal-optimized runs perform better only after having more than 22CX gates.

\begin{figure}
  \centering
  \includegraphics[width=1.0\linewidth]{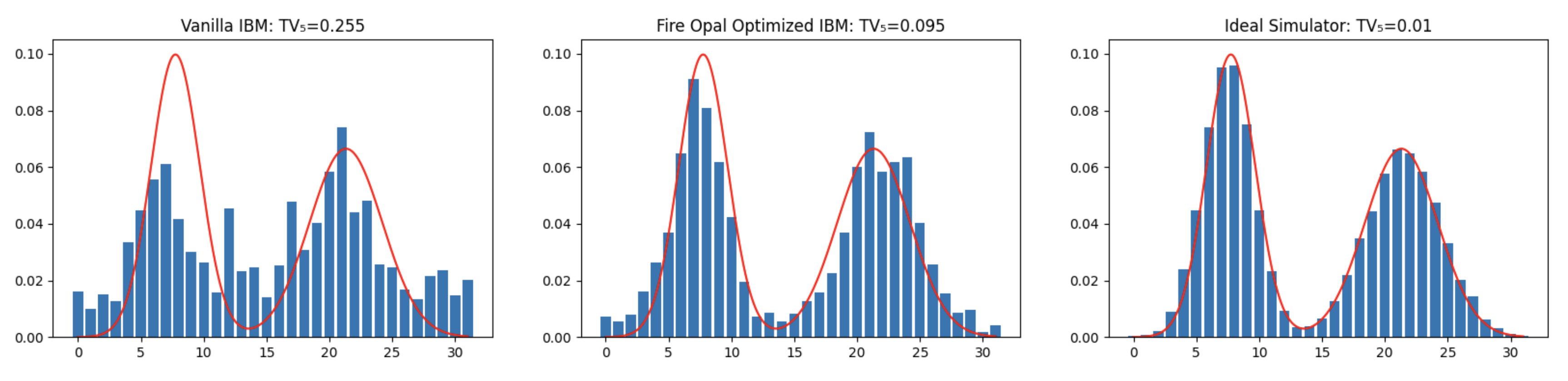}
  \caption{3 visualizations of the same 16 qubit, 38cx circuit. The red line is the underlying true distribution. We see how the accuracy of the visualization gets better from left to right, as $TV_5$ decreases. The quantum device used here is the 27 qubit IBM Algiers}\label{fig:firopal_vs_ibm_algiers_vis}
\end{figure}

To illustrate the quality of approximations for these 3 different modes we can look at figure \ref{fig:firopal_vs_ibm_algiers_vis} where we visulalize the amplitudes of the 5 most significant qubits.

% \subsection{Learning from Finance Data}

\section{Discussion}

In this article, we introduced a novel {\it hierarchical learning} method tailored for training variational circuits, focusing specifically on Quantum Circuit Born Machine models. Our method progressively learns at different scales, starting with a small number of qubits and methodically expanding to include more complex and extensive quantum circuits. Additionally, we implemented the adjoint derivative technique, which significantly accelerated derivative computations of the KL divergence, resulting in a quadratic speed-up. This combination of methods enabled us to effectively train QCBMs with 27 qubits and 1000 parameters on GPU machines, achieving an impressive $4\%$ accuracy in TV distance for approximating multivariate Gaussian distributions.

Looking ahead, we are optimistic about scaling this methodology to even larger qubit counts and circuit parameters, as our current results indicate no fundamental obstacles. Our approach appears to successfully navigate the challenges posed by Barren Plateaus in the optimization landscape of variational quantum circuits, thanks to the strategic deployment of hierarchical learning. We anticipate that our approach will be adept at learning more complex multidimensional distributions, going beyond the multidimensional and bimodal Gaussian examples discussed in the paper. In the near future, we intend to test these ideas on financial datasets to further explore and validate our methodology's capabilities.

While our focus has been on QCBMs, we anticipate that this method will be applicable to a broader spectrum of variational quantum circuits. One particularly promising direction is the application of our method to the training of variational quantum eigensolvers, which approximate the ground state of a target Hamiltonian. We envisage adapting the hierarchical learning to mirror the renormalization group (RG) flow of the Hamiltonian. This could potentially enable us to learn gate parameters at different energy scales, analogous to how we handled the digits in our QCBM case study. This expansion into new applications could significantly advance the field of quantum computing, offering more efficient and scalable solutions for complex quantum systems.

\section*{Acknowledgments}
We thank Pranav Mundada and Yulun Wang of the Q-CTRL team for their instrumental role in the effective deployment of Fire Opal optimization on the IBM quantum hardware presented in Section~\ref{sec:IBM-Fire}. We also thank Bill Fefferman, Elton Zhu, Leo Zhou, and Hovnatan Karapetyan for many fruitful discussions on the topic of variational quantum circuits and distribution loading. 

\bibliographystyle{unsrt} 
\bibliography{ref} 
%\pagebreak
\appendix
\newpage
\section{Visualizing Shallow Circuit Approximation}
\begin{figure}[H]
  \centering
  \includegraphics[width=1.0\linewidth]{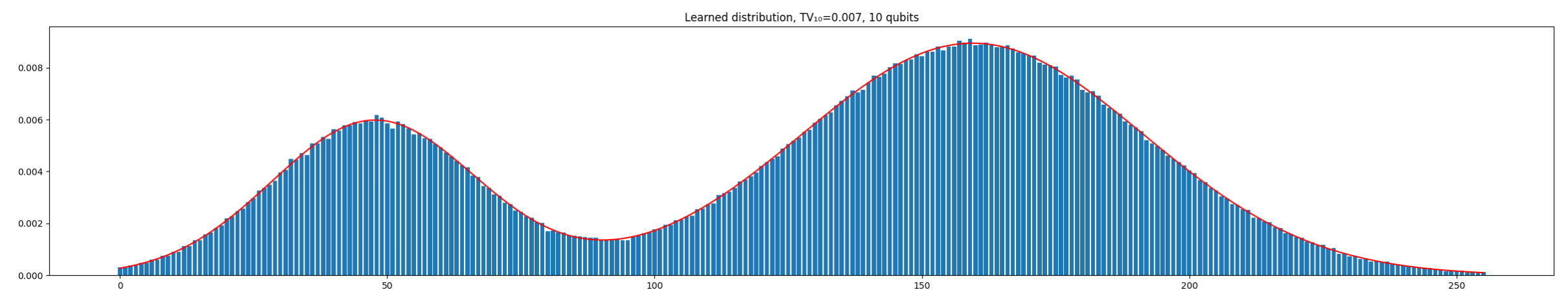}
  
  \caption{Visualizing the amplitudes of the 8 most significant qubits of the 10-qubit circuit. The underlying red line is the true distribution we are learning.}\label{fig:1d_bimodal_high_res}
\end{figure}

To illustrate the quality of our hierchical learning approximation in this and next section we will provide some visualizations of the approximated distributions. In figure \ref{fig:1d_bimodal_high_res} we see a visualization of the 8 most significant qubit amplitudes after using hierarchical learning on a bimodal Gaussian mix distribution. Here we use the grid connectivity from Fig.~\ref{fig:hierarchical} and only CX gates as 2-qubit entangling gates in our ansatz. The resulting shallow circuit only has 10 qubits, 107 single qubit gates, 48 CX gates and achieves $TV_{10}=0.007$.

\newpage
 \section{All-to-All Connectivity and 3D distribution Loading}
\begin{figure} [H]
  \centering
   \includegraphics[width=1.0\linewidth]{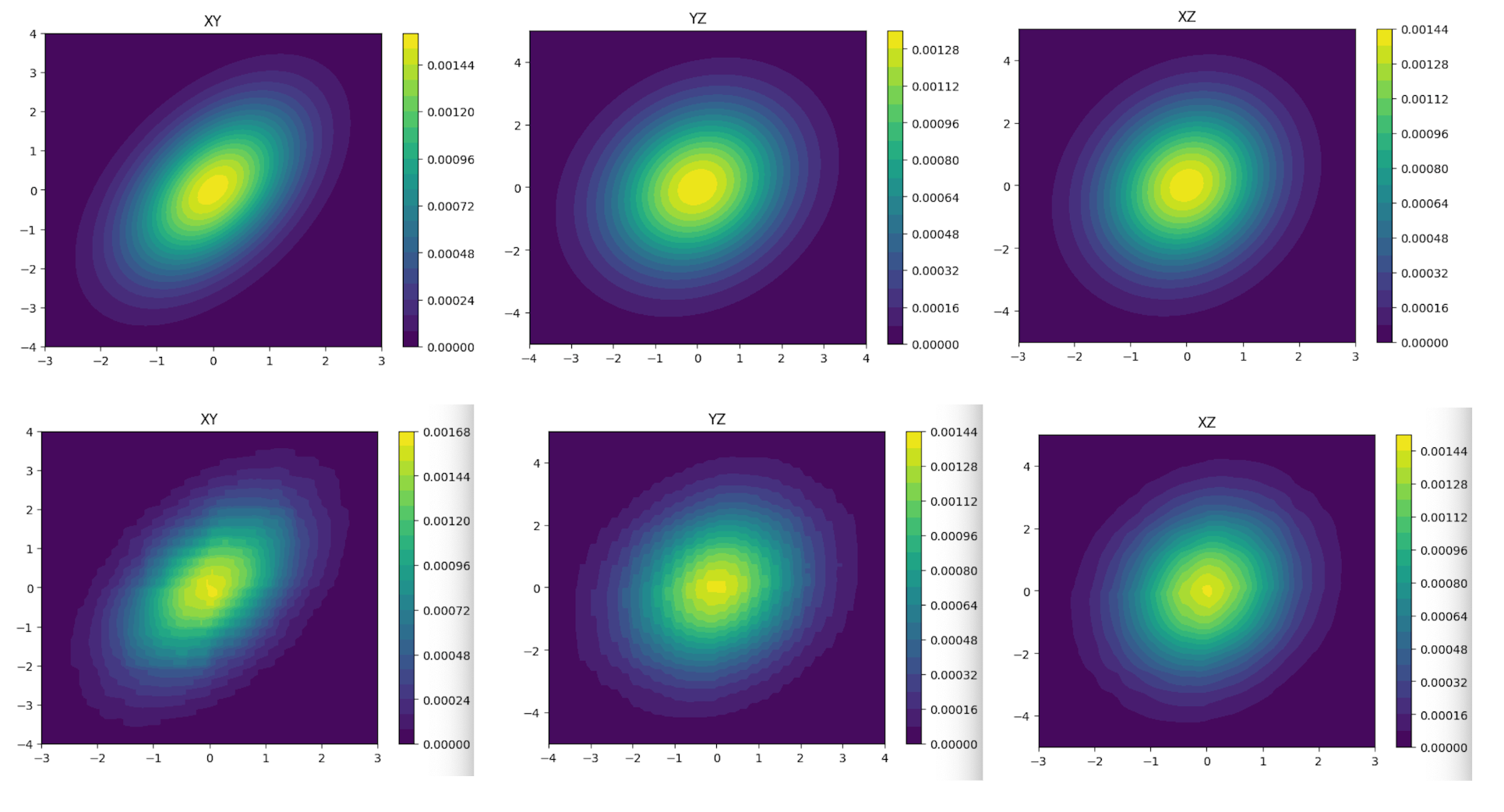}

  \caption{Comparison of the marginal distributions of the ideal vs loaded 3d multivariate normal distribution. The right column are the ideal marginals and the left column are the loaded ones.}\label{fig:3d_all_to_all}
\end{figure}

We have experimented with all-to-all connectivity and trained circuits with an ansatz that has two-qubit gates between all pairs of qubits. While this is not very practical as it results in high number of two-qubit gates even for few layers, we have noticed how good such circuits are at approximating multivariate distributions.
In Fig.~\ref{fig:3d_all_to_all} we see the side-to-side comparison of the ideal vs loaded 3d multivariate normal distribution on 18 qubits, e.g. 6 qubits per dimension. It's worth noting that this is a different 3d multivariate normal than the one presented in Fig.~\ref{fig:3d_marginals}.   
With only 2 ansatz layers we have achieved $TV_{18}=0.0285$. It's worth mentioning that there is a benefit to running such training experiments with as many seeds as possible, as unlike regular AI and ML use cases, here we do want to "overfit" our data - since we are not dealing with an inference problem but rather a data loading problem. For this particular example we ran 100 seeds.

\newpage
\section{Total Variation Resolution}\label{app:tvd}

In this section, we expand on the need to consider $\TV_n$. In particular, we use the case of a univariate Gaussian QCBM to illustrate the role that $TV_n$ plays for varying qubit size. For a single variable, we find that 10 qubits is about the cutoff beyond which increased resolution does not lead to a better $\TV$ distance on the full distribution. For smaller qubit number, we see the dramatic separation between the \TV distance on the active qubits versus the full distribution. 

We train a QCBM to learn a univariate Gaussian on 24 qubits. By tossing away one qubit at a time, we get a probability distribution on the remaining qubits. We then compare two $\TV$ distances. One is $\TV_n(p, q'_\theta)$ and the other is $\TV_{24}(p, q'_\theta)$ where $q'$ denotes the QCBM generated distribution after discarding some of the qubits. 

% \begin{center}
% \begin{tabular}{|c| c| c|} 
%  \hline
%  n & $\TV_n$ & $\TV_{24}$ \\  
%  \hline
% 24& 0.0120198566 & 0.0120198566 \\
% 23& 0.0120198566 & 0.0120198566 \\
% 22& 0.0120198566 & 0.0120198566 \\
% 21& 0.0120198566 & 0.0120198566 \\
% 20& 0.0120198566 & 0.0120198566 \\
% 19& 0.0120198564 & 0.0120198566 \\
% 18& 0.0120198560 & 0.0120198566 \\
% 17& 0.0120198542 & 0.0120198567 \\
% 16& 0.0120198416 & 0.0120198507 \\
% 15& 0.0120198067 & 0.0120198474 \\
% 14& 0.0120197005 & 0.0120198495 \\
% 13& 0.0120192226 & 0.0120197994 \\
% 12& 0.0120175400 & 0.0120197420 \\
% 11& 0.0120088738 & 0.0120195044 \\
% 10& 0.0119835773 & 0.0120193804 \\
%  9& 0.0118597269 & 0.0120212629 \\
%  8& 0.0115420650 & 0.0120533582 \\
%  7& 0.0111921801 & 0.0133957509 \\
%  6& 0.0108953766 & 0.0182164400 \\
%  5& 0.0046305578 & 0.0294763364 \\
%  4& 0.0021116980 & 0.0577599102 \\
%  3& 0.0004548054 & 0.1166881294 \\
%  2& 0.0002632882 & 0.2197497456 \\
%  1& 0.0001509644 & 0.4234306181 \\
%  \hline
% \end{tabular}
% \end{center}
\begin{figure}[H]
  \centering
  \includegraphics[width=0.8\linewidth]{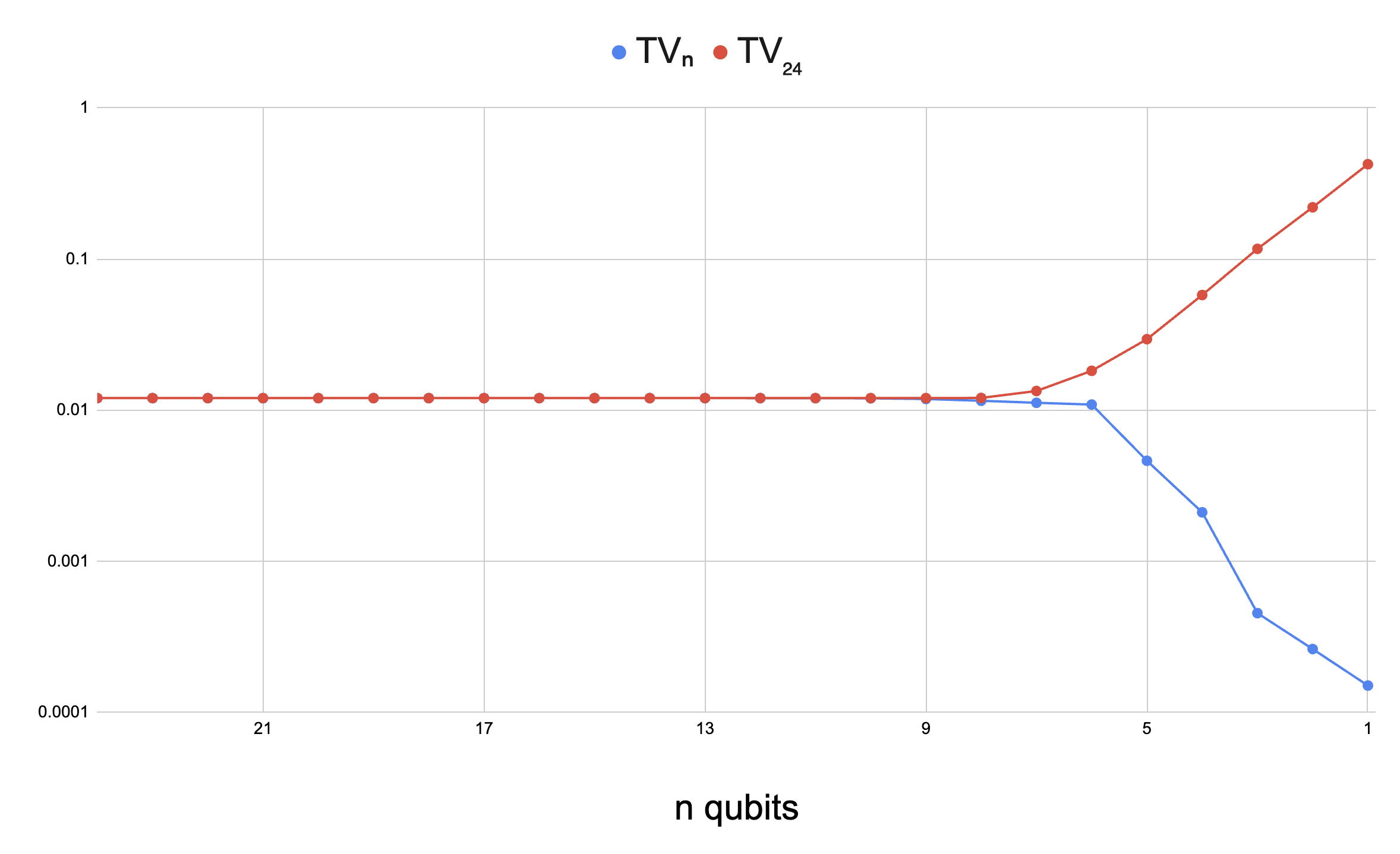}
  \caption{Given a QCBM on 24 qubits for a univariate distribution, we coarse grain the distribution by marginalizing over the least significant bits. We then compare the $\TV_n$ distance with the full $\TV_{24}$ distance. Starting at around 10 qubits we see a divergence of the two distances. In particular, we see how with relatively few qubits, one can drive $\TV_n$ down artificially.}\label{fig:tvn_comparison}
\end{figure}

\newpage

% \section{Preparing more complicated distributions}

% VQC for preparing distributions, in finance, insurance, derivative pricing.
% List of distributions
% \begin{itemize}
%     \item Log-normal, European options
%     \item Normal and exponential dist
%     \item Combination of distributions
%     \item Sophisticated high dimensional distributions
% \end{itemize}

% Error analysis and discarding qubits.

% \section{QML}
% How to use circuits for classification problems?
% Identify interesting use-case related to the finance.

\end{document}